\RequirePackage{fix-cm}
\documentclass[twocolumn,epjc3,showpacs,preprintnumbers,amsmath,amssymb]{svjour3}
\smartqed  
\RequirePackage{graphicx}
\RequirePackage[numbers,sort&compress]{natbib}
\usepackage{graphicx,amsmath,amsfonts,latexsym,amssymb,graphics,epsfig,subfigure,color,makeidx,hyperref}
\usepackage{multirow}
\usepackage{ulem}

\journalname{Eur. Phys. J. C}

\begin{document}

\title{Motion deviation of test body induced by spin and cosmological constant in extreme mass ratio inspiral binary system}

\author{Yu-Peng Zhang\thanksref{e1,addr1}
       \and Shao-Wen Wei\thanksref{e2,addr1}
       \and Pau Amaro-Seoane\thanksref{e3,addr3,addr4,addr5}
       \and Jie Yang\thanksref{e4,addr1}
       \and Yu-Xiao Liu\thanksref{e5,addr1,addr2}
}

\thankstext{e1}{e-mail:zhangyupeng14@lzu.edu.cn}
\thankstext{e2}{e-mail:weishw@lzu.edu.cn}
\thankstext{e3}{e-mail:pau@ice.cat}
\thankstext{e4}{e-mail:yangjiev@lzu.edu.cn}
\thankstext{e5}{e-mail:liuyx@lzu.edu.cn, Corresponding author}

\institute{Institute of Theoretical Physics \& Research Center of Gravitation, Lanzhou University, Lanzhou 730000, China \label{addr1}
\and
Key Laboratory for Magnetism and Magnetic of the Ministry of Education, Lanzhou University, Lanzhou 730000, China \label{addr2}
\and
Institute of Space Sciences (ICE, CSIC) \& Institut d'Estudis Espacials de Catalunya (IEEC)
at Campus UAB, Carrer de Can Magrans s/n 08193 Barcelona, Spain \label{addr3}
\and
Institute of Applied Mathematics, Academy of Mathematics and Systems Science, CAS, Beijing 100190, China \label{addr4}
\and
Kavli Institute for Astronomy and Astrophysics, Beijing 100871, China \label{addr5} }

 \maketitle

\begin{abstract}
The future space-borne detectors will provide the possibility to detect gravitational waves emitted
from extreme mass ratio inspirals of stellar-mass compact objects into supermassive black holes. It is natural to expect that the spin of the compact object and cosmological constant will affect the orbit of the inspiral process and hence lead to the considerable phase shift of the corresponding gravitational waves. In this paper, we investigate the motion of a spinning test particle in the spinning black hole background with a cosmological constant and give the order of motion deviation induced by the particle's spin and the cosmological constant by considering the corresponding innermost stable circular orbit. By taking the neutron star or kerr black hole as the small body, the deviations of the innermost stable circular orbit parameters induced by the particle's spin and cosmological constant are given. Our results show that the deviation induced by particle's spin is much larger than that induced by cosmological constant when the test particle locates not very far away from the black hole, the accumulation of phase shift during the inspiral from the cosmological constant can be ignored when compared to the one induced by the particle's spin. However when the test particle locates very far away from the black hole, the impact from the cosmological constant will increase dramatically. Therefore the accumulation of phase shift for the whole process of inspiral induced by the cosmological constant and the particle's spin should be handled with caution.\\

\end{abstract}

\maketitle

\section{Introduction}\label{scheme1}

Gravitational waves have been directly detected by the Laser Interferometer
Gravitational-Wave Observatory (LIGO) and Virgo from merging black holes and
inspiraling neutron star
binaries\cite{Abbott2016a}.
These systems have been detected with a mass ratio of the order of $1$ (as predicted by
\cite{Amaro-SeoaneChen2016}, who also predicted that these binaries should
predominately have low spin values and essentially be circular). Since
intermediate-mass ratio inspirals are already detectable by ground-based
detectors, see \cite{Amaro-Seoane2018a}, we will have to wait for space-borne
detectors such as the Laser Interferometer Space Antenna (LISA)
\cite{lisapaper,lisawebsite}, DECIGO\cite{Shuichi2017a,Shuichi2017b}, Taiji \cite{taiji,taijisource}, and
Tianqin \cite{Luo2016} to detect extreme-mass ratio inspirals, i.e. the progressive
inspiral of a stellar-mass compact object on to a supermassive black hole, see
\cite{Amaro-Seoane2018b}.

For an extreme-mass ratio inspiral system (EMRI), a test particle with the motion along geodesic is the simplest description for the small body.
While a small body always possesses spin angular momentum in such system, in order to describe the EMRI system more accurately,
the motion of the small body can be envisaged as a spinning test body inspiraling into a supermassive black hole.
A test particle without spin can be treated as a point-like particle and its
motion in a curved spacetime is described by geodesics. However, when the
reaction of the test particle is considered, the motion does not comply
with a geodesic \cite{Warburton:2011hp,Isoyama:2014mja,vandeMeent:2016hel}.
Likewise, the motion of a spinning test particle does not follow a geodesic
because of the additional spin-curvature force \cite{Hanson1974,Wald:1972sz}.
As the descriptions for the spinning test particle, the spin of the test particle indeed make contributions to its motion and the corresponding spin should be considered.

Like the spin of a test particle, a non-zero cosmological constant also affects the motion of the test particle.
The observations \cite{Riess_1998,Perlmutter_1999} have shown that the cosmological constant is positive and non-zero
with a confidence of $P(\Lambda>0)=99\%$. Therefore contributions to the motion of the particle in the black hole
background from the non-zero cosmological constant should be considered.
The cosmological constant was firstly proposed by Einstein in order to obtain a
static universe. A black hole with a negative cosmological constant will have a thermodynamic behavior, and
there exists a phase transition between the stable large black hole and the thermal gas phase
\cite{Hawking}. The AdS/CFT correspondence and Holography has been addressed by the works of
\cite{Maldacena,Witten}, and the small-large black hole phase transition in the
charged or rotating AdS black hole backgrounds was also investigated in Refs. \cite{Chamblin,Caldarelli,Weiliu}.

We should note that the magnitudes of the particle's spin and current observed cosmological constant are
very tiny and the corresponding motion deviations caused by them can be ignored.
That's why most of the works about the EMRI are always described by a test particle with the motion along the geodesic trajectories in a black hole background.
While the key thing should be noted here is that the deviation of the orbital angular frequency induced by the
particle's spin and cosmological constant will accumulate and lead to considerable phase shift during the
long time inspiraling process of the small body in the EMRI system, this phenomena was also reported in
Ref. \cite{Han2010}. Therefore, if we want to investigate the motion of a test particle more accurately
in the EMRI system, the relationship between particle motion and both particle's spin
and cosmological constant should be clarified.

Circular orbits below the innermost stable circular orbit (ISCO) are unstable, and it is always regarded as the beginning
of the merger of the binary. The motion behavior of particles in ISCO is closely related to the nature of black holes and we will use it to investigate the motion deviation induced by the spin of test particle and the
non-zero cosmological constant. The ISCO of a spinning test particle
in the Schwarzschild, Kerr, and Kerr-Newman (KN) black hole backgrounds have been investigated
in~\cite{Suzuki:1997by,Jefremov:2015gza,zhang2018:zwgsl}. Equatorial circular orbits and ISCOs in different black hole
backgrounds have been investigated systematically in the related literature
(see Refs.~\cite{Pugliese:2013zma,Kaplan,Landau,Suzuki:1997by,Shibata:1998ih,Zdunik:2000qn,Baumgarte:2001ab,Grandclement:2001ed,Miller:2003vc,Marronetti:2003hx,GondekRosinska:2004eh,Campanelli:2006gf,Shahrear:2007zz,Cabanac:2009yz,Abdujabbarov:2009az,Hadar:2011vj,Akcay:2012ea,Hod:2013vi,Chakraborty:2013kza,Hod:2014tpa,Isoyama:2014mja,Zahrani:2014rqa,Zaslavskii:2014mqa,Delsate:2015ina,Chartas:2016ckd,Harms:2016ctx,Lukes-Gerakopoulos:2017vkj,Jai-akson:2017ldo,zhang2018:zwgsl,ChenYuLiu2018,Luk:2018xmt,Sajal2018,Warburton:2015hp,Hyun-Chu2017,Zhang:2018eau}).
The motion of the spinning test particle in the Horndeski theory was also investigated in \cite{Sajal20182}. Some works have addressed the motion of a spinning test
particle with non-zero cosmological constant in Schwarzschild and Kerr-de Sitter spacetimes
\cite{stuchlik_1999,stuchlik_2006,Plyatsko_2017,Plyatsko_2018}, for which the equilibrium conditions and nonequatorial circular orbits were investigated.

In this paper, we will investigate the motion deviation induced by the spin of a test particle and the non-zero cosmological constant by considering the motion of a spinning test particle in a rotating black hole with zero cosmological constant.
Firstly, we review the equations of motion for a spinning test particle in curved spacetime and derive the corresponding four-momentum and tangent vector along the trajectory in a Kerr-dS/AdS black hole background in Sec.~\ref{scheme1}. By setting the particle's spin and cosmological constant to zero, we can get the original no-deviation geodesic motion of the test particle in Kerr black hole background, the motion deviation can be obtained by comparing the results that for the original case and non-zero particle's spin and cosmological constant case. In Sec.~\ref{scheme2}, we derive the ISCO for the spinning test particle in Kerr-dS/AdS black hole, with these results in hand, the motion of small body with non-zero spin and cosmological constant can be obtained. Thanks to the real magnitudes of the particle's spin and cosmological constant are very tiny and we can simplify our results in the linear order approach of particle's spin and cosmological constant. Then we have the analytic angular frequency with the linear order approach, which is used to naively estimate the corresponding magnitude of the phase shift induced by them. Finally, a brief summary and conclusion are given in Sec.~\ref{Conclusion}.

\section{Motion of a small body in black hole background }{\label{scheme1}}

In this section, we solve the equations of motion of a spinning test particle in a Kerr-dS/AdS black hole background, where a unified treatment between the cases of geodesics in a Kerr-dS/AdS black hole and spinning test particle in a Kerr-dS/AdS black hole is presented. We will use the ``pole-dipole'' approximation to describe the motion of the spinning test particle. The corresponding equations of motion for the spinning test
particle, also known as the Mathisson-Papapetrou-Dixon (MPD) equations, can be
found in Refs.~\cite{Mathisson,Papapetrou1951a,Papapetrou1951c,Dixon,phdthesis,Hojman1977,Bahram2006,Zalaquett2014,Ruangsri:2015cvg}.
The four-velocity $u^\mu$ and the four-momentum $P^\mu$ in this
scenario are not parallel \cite{phdthesis,Armaza2016,Hojman2013}. The
four-momentum keeps timelike along the trajectory and satisfies $P^\mu
P_\mu=-m^2$ with $m$ the mass of the test particle, while the four-velocity
might be superluminal \cite{phdthesis,Armaza2016,Hojman2013} if the spin of the
test particle is too large. When the multi-pole effects are
considered, the superluminal problem can be avoided
\cite{Deriglazov:2015zta,Deriglazov:2015wde,Ramirez:2017pmp,Deriglazov:2017jub,Jan2010}.
The gravitational radiation of a spinning test particle has been calculated
by using the perturbation method, see Refs.
\cite{Lukes-Gerakopoulos:2017vkj,Han2008,Han2010,Warburton2017}. The
collisional Penrose process with spinning particles were also investigated in Refs.
\cite{Liu:2018myg,Mukherjee:2018kju}.

The MPD equations can be derived with several methods, such as the multi-pole expansion \cite{Papapetrou1951a,Papapetrou1951c,Mathisson,Dixon,Faye:2006gx}, the lagrangian method  \cite{phdthesis,Hojman2013}, and the hamiltonian method \cite{Tauber1988,Lukes-Gerakopoulos2019,Lukes-Gerakopoulos2014}. The corresponding forms with the pole-dipole approximations read
    \begin{eqnarray}
    \frac{D P^{\mu}}{D \lambda} &=& -\frac{1}{2}R^\mu_{\nu\alpha\beta}u^\nu S^{\alpha\beta},\label{equationmotion1}\\
    \frac{D S^{\mu\nu}}{D \lambda} &=&P^\mu u^\nu-u^\mu P^\nu, \label{equationmotion2}
    \end{eqnarray}
where $P^{\mu}$, $S^{\mu\nu}$, and $u^{\mu}$ are the four-momentum, spin tensor, and tangent vector of the spinning test particle along the trajectory, respectively. Note that, the $\frac{D}{D\lambda}$ is a covariant derivative along the trajectory of the particle, and $\lambda$ is the affine parameter of the test particle. Obviously, the motion of the particle does not follow the geodesic due to the spin-curvature force $-\frac{1}{2}R^\mu_{\nu\alpha\beta}u^\nu S^{\alpha\beta}$.

We should use a ``spin-supplementary condition'' to determine the motion of a spinning test particle since the motion described by the MPD equations is not uniquely specified. The choice of this condition is not unique because it is related to the center of mass of the spinning test particle with different observers \cite{Lukes-Gerakopoulos:2017vkj,Wald:1972sz,Lukes-Gerakopoulos2014,Costa2017}. In this paper we adopt the Tulczyjew spin-supplementary condition \cite{Tulczyjew}:
\begin{equation}
P_\mu S^{\mu\nu}=0,
\label{supplementarycondition}
\end{equation}
for which the four-momentum $P^\mu$ satisfies
\begin{equation}
P^\mu P_\mu=-m^2,\label{normal1}
\end{equation}
and it keeps timelike along the trajectory, while the tangent vector velocity $u^\mu$ may transform from timelike to spacelike as it is not parallel to $P^\mu$ \cite{phdthesis,Armaza2016,Hojman2013}.

The Kerr-dS/AdS black hole background can be described by the following metric in the Boyer-Lindquist coordinates
    \begin{eqnarray}
    ds^2\!\!&=&\!\!-\frac{\Delta}{\rho^2}\left(dt-\frac{a\sin^2\theta}{1+\frac{\Lambda}{3}a^2} d\phi\right)^2
        +\frac{\rho^2 d\theta^2}{1+\frac{\Lambda}{3}a^2\cos^2\theta } \nonumber\\
      \!\!&+&\!\!\frac{(1+\frac{\Lambda}{3}a^2\cos^2\theta)\sin^2\theta}{\rho^2}
            \left(\frac{r^2+a^2}{1+\frac{\Lambda}{3}a^2}d\phi-adt\right)^2\nonumber\\
        \!\!&+&\!\!\frac{\rho^2}{\Delta}dr^2,
      \label{metric}
    \end{eqnarray}
where the metric functions $\Delta$ and $\rho^2$ are
\begin{eqnarray}
\Delta &=&  (r^2+a^2)\left(1-\frac{\Lambda}{3}r^2\right)-2M r, ~~\\
 \rho^2 &=& r^2+a^2\cos^2\theta.
\end{eqnarray}
Here $\Lambda$, $M$, and $a=\frac{J}{M}$ are the cosmological constant, mass, and spin of the black hole, respectively. We set the gravitational constant $G=1$ and the speed of light $c=1$.

In this paper, we only consider the equatorial motion of the spinning test particle with the spin-aligned or anti-aligned orbits, the four-momentum and spin tensor satisfy $P^\theta=0$ and $S^{\theta \mu}=0$. The non-vanishing independent variables for the equatorial orbits are $P^t$, $P^r$, $P^\phi$, and $S^{r\phi}$. By using the spin-supplementary condition (\ref{supplementarycondition}), the remaining components of spin tensor are \cite{Hojman1977}
\begin{eqnarray}
S^{r t}=-S^{r\phi}\frac{P_{\phi}}{P_t}, ~~~~S^{\phi t}=S^{r\phi}\frac{P_r}{P_t}.
\label{spintensor}
\end{eqnarray}
Substituting Eq. (\ref{spintensor}) into following equation
\begin{equation}
s^2=\frac{1}{2}S^{\mu\nu}S_{\mu\nu}=S^{\phi r}S_{\phi r}+S^{t r}S_{t r}+S^{t\phi}S_{t\phi},
\end{equation}
and by using Eq. (\ref{normal1}), the $r-\phi$ component of the spin tensor reads
\begin{equation}
S^{r\phi}=-\frac{s}{r}\frac{P_{t}}{m}\left|\frac{a^2\bar{\Lambda}+3M^2}{3M^2}\right|=-\frac{s}{r}\frac{P_{t}}{m}\left|h\right|,\nonumber\\
\end{equation}
where the parameter $h=\left|\frac{a^2\bar{\Lambda}+3M^2}{3M^2}\right|$ and the dimensionless parameter $\bar{\Lambda}=\Lambda~M^2$.  Finally we obtain the non-vanishing components of the spin tensor $S^{\mu\nu}$ in the Kerr-dS/AdS black hole background:
\begin{eqnarray}
S^{r\phi}&=&-S^{\phi r}=-\frac{s}{r}\frac{P_{t}}{m}h,\nonumber\\
S^{rt}&=&-S^{tr}=-S^{r\phi}\frac{P_\phi}{P_t}=\frac{s}{r}\frac{P_\phi}{m}h,\label{spinnozo}\\
S^{\phi t}&=&-S^{t\phi}=S^{r\phi}\frac{P_r}{P_t}=-\frac{s}{r}\frac{P_r}{m}h,\nonumber
\end{eqnarray}
where the parameter $s$ is the spin angular momentum of the test particle and the spin perpendicular to the equatorial plane. With the non-vanishing spin tensor, the non-zero spatial component of the spin angular momentum is \cite{phdthesis}
\begin{equation}
S_z=rS^{r\phi}=-s\frac{P_t}{m},
\end{equation}
where the index $z$ stands for the direction of the spin for the test particle.

Compared to the case of a particle without spin, the conserved quantities have changed due to the spin of the test particle. For a killing vector field $\mathcal{K}^\mu$, the corresponding conserved quantity is \cite{Hojman1977,phdthesis}
\begin{equation}
\mathcal{C}=\mathcal{K}^\mu P_\mu-\dfrac{1}{2} S^{\mu \nu}\mathcal{K}_{\mu;\nu},
\label{Eq:Conserved_quantity}
\end{equation}
where the semicolon denotes the covariant derivative. Since there are a timelike Killing vector $\xi^\mu=(\partial_t)^\mu$ and a spacelike Killing vector $\eta^\mu=(\partial_\phi)^\mu$
in the spacetime (\ref{metric}), we have two conserved quantities \cite{Hojman1977}
\begin{eqnarray}
e&=&-\mathcal{C}_t=-\xi^\mu P_\mu+\dfrac{1}{2} S^{\mu \nu}{\xi}_{\mu;\nu}\nonumber\\
&=&-P_t+\frac{1}{2}\frac{s}{r}\frac{P_t}{m}h\partial_rg_{t\phi}-\frac{1}{2}\frac{s}{r}\frac{P_\phi}{m}h\partial_r g_{tt},\label{conservedenergy}\\
j&=&\mathcal{C}_{\phi}=\eta^\mu P_\mu-\dfrac{1}{2} S^{\mu \nu}{\eta}_{\mu;\nu}\nonumber\\
&=&P_\phi+\frac{1}{2}\frac{s}{r}\frac{P_\phi}{m}h\partial_rg_{\phi t}-\frac{1}{2}\frac{s}{r}\frac{P_t}{m}h\partial_r g_{\phi\phi},\label{conservedmomentum}
\end{eqnarray}
where $e$ and $j$ are the energy and total angular momentum of the spinning test particle. One can verify the relations
$S^{\mu \nu}{\xi}_{\mu;\nu}=S^{\mu\nu}\xi^\beta\partial_\nu g_{\beta \mu}$ and $S^{\mu \nu}{\eta}_{\mu;\nu}=S^{\mu\nu}\eta^\beta\partial_\nu g_{\beta \mu}$ for the two Killing vectors.

By solving Eqs. \eqref{normal1}, \eqref{conservedenergy}, and \eqref{conservedmomentum}, the non-vanishing  components of the four-momentum are

\begin{eqnarray}
    P_t&=&-\frac{2 m r \left(\bar{e}\bar{s}M h \partial_rg_{\phi t}+2 \bar{e}  r+h \bar{s}\bar{j}M^2 \partial_r g_{t t}\right)}{4 r^2+h^2 \bar{s}^2M^2 \big(\partial_r g_{\phi\phi} \partial_rg_{t t}-(\partial_rg_{\phi t})^2\big)}\nonumber\\
    &=&\frac{m}{3\Gamma}\Bigg\{\bar{j}M^2\bar{s}(3+a^2\frac{\bar{\Lambda}}{M^2})(3M-r^3\frac{\bar{\Lambda}}{M^2})\nonumber\\
    &&+3\bar{e}r^3(aM\bar{s}\frac{\bar{\Lambda}}{M^2}-3)-9a\bar{e}M^2\bar{s}\bigg\} , \label{memantumpt}
\end{eqnarray}
\begin{eqnarray}
    P_\phi&=&\frac{2 m r \left(2 \bar{j}M r-\bar{e}\bar{s} M h \partial_rg_{\phi\phi}-h \bar{j}\bar{s}M^2 \partial_r g_{\phi t}\right)}{h^2 \bar{s}^2M^2 \left(\partial_rg_{\phi\phi} \partial_rg_{t t}-(\partial_rg_{\phi t})^2\right)+4  r^2}\nonumber\\
    &=&\frac{mM}{\Gamma\left(3+a^2\frac{\bar{\Lambda}}{M^2}\right)}\bigg\{9a^2\bar{e}M\bar{s}+\left(3+a^2\frac{\bar{\Lambda}}{M^2}\right)\nonumber\\
    &&\bigg(\bar{j}r^3\bigg(3+aM\bar{s}\frac{\bar{\Lambda}}{M^2}\bigg)-3a\bar{j}M^2\bar{s}-3\bar{e}r^3\bar{s}\bigg)\bigg\},\label{memantumpp}
\end{eqnarray}
and
\begin{eqnarray}
    (P^r)^2 &=&-\frac{m^2+g^{\phi\phi}P_\phi^2+2g^{\phi t}P_\phi P_t+g^{tt}P_t^2}{g_{rr}},  \label{memantumpr}
\end{eqnarray}
where the function $\Gamma$ is
\begin{equation}
\Gamma=r^3(3+\bar{\Lambda}\bar{s}^2)-3M^3\bar{s}^2,
\label{gamma}
\end{equation}
and the dimensionless parameters are defined as $\bar{e}=\frac{e}{m}$, $\bar{j}=\frac{j}{mM}$, and $\bar{s}=\frac{s}{mM}$, respectively.

Because the trajectories of the test particle are independent of the affine parameter $\lambda$ \cite{Dixon,Georgios2017}, in this paper we take the affine parameter $\lambda$ as coordinate time and set $u^t=1$. With this choice, the orbital frequency reads
\begin{equation}
\Omega=\dot{\phi}.
\end{equation}
By substituting $u^t=1$ and the non-zero components of $S^{\mu\nu}$ (\ref{spinnozo}) into the equations of motion (\ref{equationmotion1}) and (\ref{equationmotion2}), we have  \cite{Hojman2013,Zhang:2016btg}
\begin{eqnarray}
\frac{DS^{tr}}{D t}&=&P^t\dot{r}-P^r\nonumber\\
        &=&\frac{h}{2m}\frac{s}{r}g_{\phi \mu}R^\mu_{\nu\alpha\beta}u^\nu S^{\alpha\beta}
           +h\frac{s}{r}\frac{P_\phi}{rm}\dot{r}, \label{spinvelocityequation1}\\
\frac{DS^{t\phi}}{D t}
        &=& P^t\dot{\phi}-P^\phi\nonumber\\
        &=& -\frac{h}{2m}\frac{s}{r}g_{r \mu}R^\mu_{\nu\alpha\beta}u^\nu S^{\alpha\beta}
            -h\frac{s}{r}\frac{P_r}{rm}\dot{r}\label{spinvelocityequation2}.
\end{eqnarray}
Then we derive the tangent vector $u^\mu$ as follows
\begin{eqnarray}
\dot{r}&=&\frac{b_2 c_1-b_1c_2}{a_2b_1-a_1b_2},\\
\dot{\phi}&=&\frac{a_2 c_1-a_1c_2}{a_1b_2-a_2b_1}.
\end{eqnarray}
Here the parameters $a_1$, $b_1$, $c_1$, $a_2$, $b_2$, and $c_2$ are
\begin{eqnarray}
a_1&=&P^t-h\frac{s}{r}\frac{P_\phi}{mr}+\frac{h}{2m}\frac{s}{r}R_{\phi r\mu\nu}S^{\nu\mu},\\
b_1&=&\frac{1}{2m}\frac{s}{r}R_{\phi \phi\mu\nu}S^{\nu\mu},\\
c_1&=&-P^r+\frac{h}{2m}\frac{s}{r}R_{\phi t\mu\nu}S^{\nu\mu},\\
a_2&=&h\frac{sP_r}{r^2}-\frac{h}{2m}\frac{s}{r}R_{r r\mu\nu}S^{\nu\mu},\\
b_2&=&P^t-\frac{h}{2m}\frac{s}{r}R_{r\phi \mu\nu}S^{\nu\mu},\\
c_2&=&-P^\phi-\frac{h}{2m}\frac{s}{r}R_{rt\mu\nu}S^{\nu\mu},
\end{eqnarray}
where
\begin{eqnarray}
R_{\phi t\mu\nu}S^{\nu\mu} &\propto& P^r,~(c_1\propto P^r),\\
R_{\phi \phi \mu\nu}S^{\nu\mu} &\propto& P^r, ~(b_1\propto P^r).
\end{eqnarray}
So the radial momentum $P^r$ and radial component of the tangent vector $u^r$ are parallel \cite{zhang2018:zwgsl,Zhang:2016btg}, and we can use the radial component of the four-momentum $P^r$ to define the effective potential of the spinning test particle. Note that, the tangent vector along the trajectory of the spinning test particle actually corresponds to the relativistic center-mass four-velocity when the $\lambda$ is the proper time $\tau$. Moreover, there is
an explicit relation giving the tangent vector as a function of the four-momentum
$P^\mu$ and the spin tensor $S^{\mu\nu}$, see details in  Refs. \cite{Georgios2017,EhlersRudolph1977}.

\section{ISCO of a spinning particle in Kerr-dS/AdS black hole background and estimation of phase shift}\label{scheme2}

We know that the motion of a test particle in a central field can be solved in terms of the radial coordinate in the Newtonian dynamics \cite{Kaplan,Landau}. And the motion of a test particle in the black hole background can also be solved by using the effective potential method in general relativity.

Due to the radial velocity $u^r$ is parallel to the radial component $P^r$, we can obtain the effective potential of the spinning test particle in the Kerr-dS/AdS black hole background by using the form of $P^r$ \eqref{memantumpr} \cite{Jefremov:2015gza}. We then have \cite{zhang2018:zwgsl,Jefremov:2015gza,misnerthornewheeler}
    \begin{eqnarray}
    \frac{(P^r)^2}{m^2}&=&\left(\alpha \bar{e}^2+\beta \bar{e}+\gamma\right)=\left(\bar{e}-\frac{-\beta+\sqrt{\beta^2-4\alpha\gamma}}{2\alpha}\right)\nonumber\\
    &\times& \left(\bar{e}-\frac{-\beta-\sqrt{\beta^2-4\alpha\gamma}}{2\alpha}\right),\label{effectivepotentiala}
    \end{eqnarray}
where the functions $\alpha$, $\beta$, and $\gamma$ are
\begin{eqnarray}
\alpha &=&\frac{1}{\Gamma^{2}} \Bigg\{a^2M^2r^6\bar{s}^2\left(\frac{\bar{\Lambda}}{M^2}\right)^2+9r^6+9a^2r^3(2M+r)\nonumber\\
&&+18M^2r(a^3+3ar)\bar{s}+9M^2\bigg[(2M-r)r^3\nonumber\\
&&+a^2M(M+2r)\bigg]\bar{s}^2+3r^3\frac{\bar{\Lambda}}{M^2}\bigg[a^2r(a^2+r^2)\nonumber\\
&&+M^2(r^3-a^2(2M+r))\bar{s}^2\bigg]
\Bigg\},
\end{eqnarray}
\begin{eqnarray}
\beta &=&-\frac{2M}{3\Gamma^2}(a^2 \frac{\bar{\Lambda}}{M^2} +3) \bar{j} \Bigg\{9M\bigg[2ar^3+(3M-r)r^3\bar{s}\nonumber\\
&&+2a^2Mr\bar{s}+aM^2(M+r)\bar{s}^2\bigg]\nonumber\\
&&+3ar^3\bigg[a^2r-aMr\bar{s}
+r^3-M^2(2M+r)\bar{s}^2
\bigg]\frac{\bar{\Lambda}}{M^2}\nonumber\\
&&+aM^2r^6\bar{s}^2\left(\frac{\bar{\Lambda}}{M^2}\right)^2\Bigg\},
\end{eqnarray}
and
\begin{eqnarray}
\gamma &=& \frac{1}{9r^2\Gamma^2}\Bigg\{3\bigg[a^2(r^2\frac{\bar{\Lambda}}{M^2}-3)+r(6M-3r+r^3\frac{\bar{\Lambda}}{M^2})\bigg]\nonumber\\
&&\bigg[r^3(3+M^2\bar{s}^2\frac{\bar{\Lambda}}{M^2})-3M^3\bar{s}^2\bigg]^2+
\bigg[9M^4\bar{s}^2\nonumber\\
&&-6M^3r^3\bar{s}^2\frac{\bar{\Lambda}}{M^2}+3r^4(a^2\frac{\bar{\Lambda}}{M^2}+r^2\frac{\bar{\Lambda}}{M^2}-3)\bigg]\nonumber\\
&&\times\bar{j}^2M^2r^2(3+a^2\frac{\bar{\Lambda}}{M^2})^2-6Mr^3(ar\bar{s}\frac{\bar{\Lambda}}{M^2}-3)\nonumber\\
&&+M^2r\bar{s}\left(18a+r^5\bar{s}\left(\frac{\bar{\Lambda}}{M^2}\right)^2\right)\Bigg\}.\label{gamma1}
\end{eqnarray}

The effective potential of the spinning test particle in the Kerr-dS/AdS black hole background is given by
\begin{equation}
V_{\text{eff}}^{\text{spin}}=\frac{-\beta+\sqrt{\beta^2-4\alpha\gamma}}{2\alpha}\label{spinningeffective},
\end{equation}
where the effective potential of the test particle is defined by the positive square root of Eq. (\ref{effectivepotentiala}), because the positive square root corresponds to the four-momentum pointing toward future, while the negative corresponds to the past-pointing four-momentum \cite{misnerthornewheeler}.

If a test particle satisfies the following two conditions \cite{Jefremov:2015gza}, it does not move in the radial direction:
\begin{equation}
\dot{r}=0~ \left(V_{\text{eff}}=0\right)\label{condition1},
\end{equation}
the acceleration in the radial direction is zero:
\begin{equation}
\ddot{r}=0~ \left(\frac{dV_{\text{eff}}}{dr}=0\right)\label{condition2},
\end{equation}
then its motion is a circular orbit.

The outer circular orbit locates at the minimum position of the effective potential is stable, and the inner one locates at the maximum position of $V_{\text{eff}}$ is unstable. The ISCO locates at the position where the maximum and minimum of the effective potential merge. Thus, for the ISCO of a test particle, the effective potential should also satisfy
\begin{equation}
\frac{d^2V_{\text{eff}}}{dr^2}=0\label{condition3}.
\end{equation}

Then, we can use Eqs. \eqref{condition1}, \eqref{condition2}, and \eqref{condition3} to derive the ISCO of a test particle. Note that, since the equation of motion of the spinning test particle is obtained by using the ``pole-dipole'' approximation, its tangent vector velocity will transform from timelike to spacelike if the particle's spin is too large. For simplicity, in this paper we neglect the higher-order ``multi-pole'' and only use Eqs. \eqref{equationmotion1} and \eqref{equationmotion2} to describe the motion of the spinning test particle with the following superluminal constraint \cite{zhang2018:zwgsl}
\begin{eqnarray}
\frac{-1}{(\frac{dt}{d\tau})^2} =
  \frac{g_{tt}}{c^2}
  + g_{rr}\Big(\frac{\dot{r}}{c}\Big)^2
  + g_{\phi\phi}\Big(\frac{\dot{\phi}}{c}\Big)^2
  + 2g_{\phi t}\frac{\dot{\phi}}{c^2}<0. \label{velocitysquare}
\end{eqnarray}
Actually, the spin of the test particle is so tiny and the corresponding motion will not exceed the speed of light.

\begin{figure*}[!htb]
	\subfigure[~$a=0$]{\includegraphics[width=0.3\textwidth]{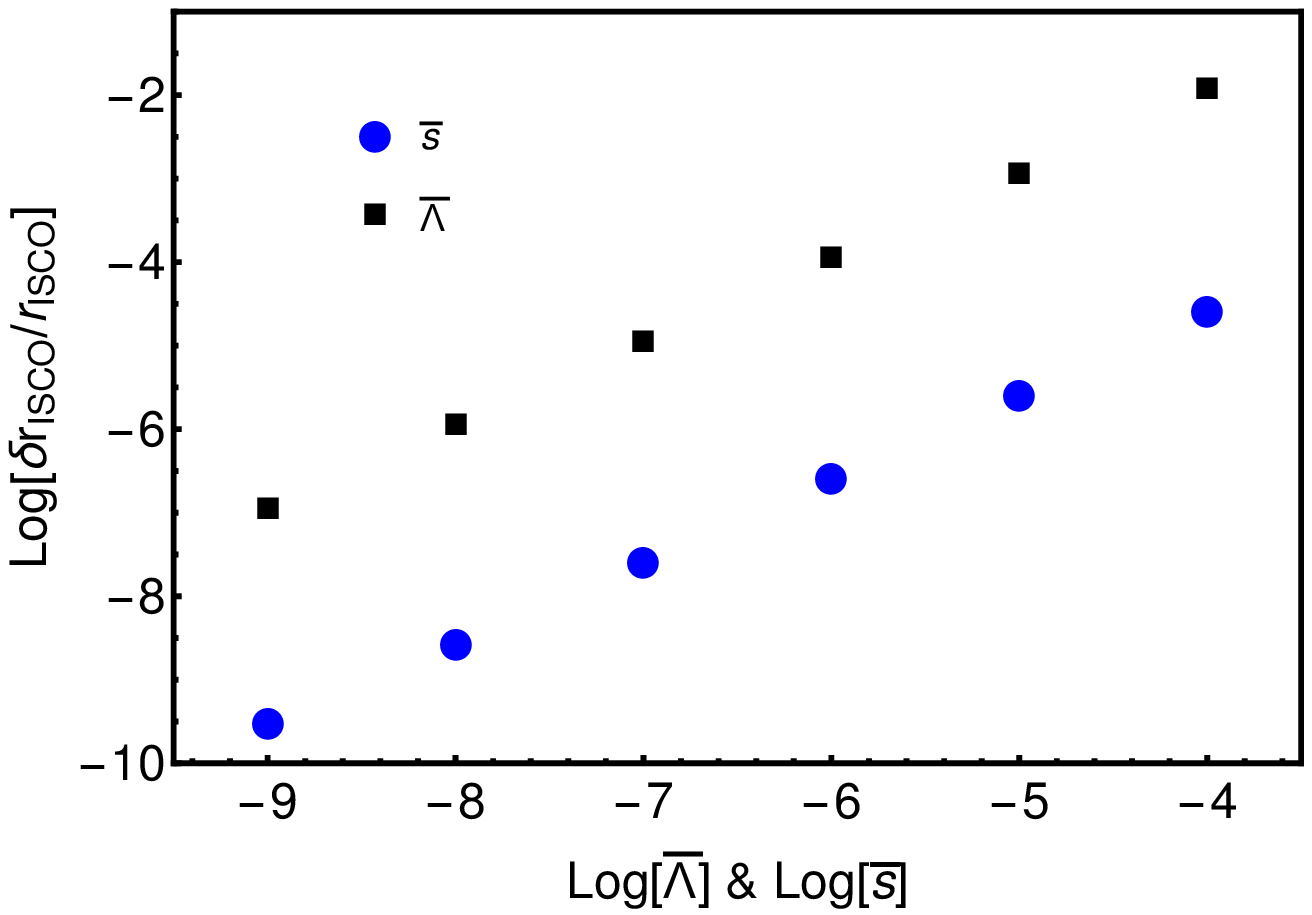}}
	\subfigure[~$a=0.25$(counter-rotating orbit)]{\includegraphics[width=0.3\textwidth]{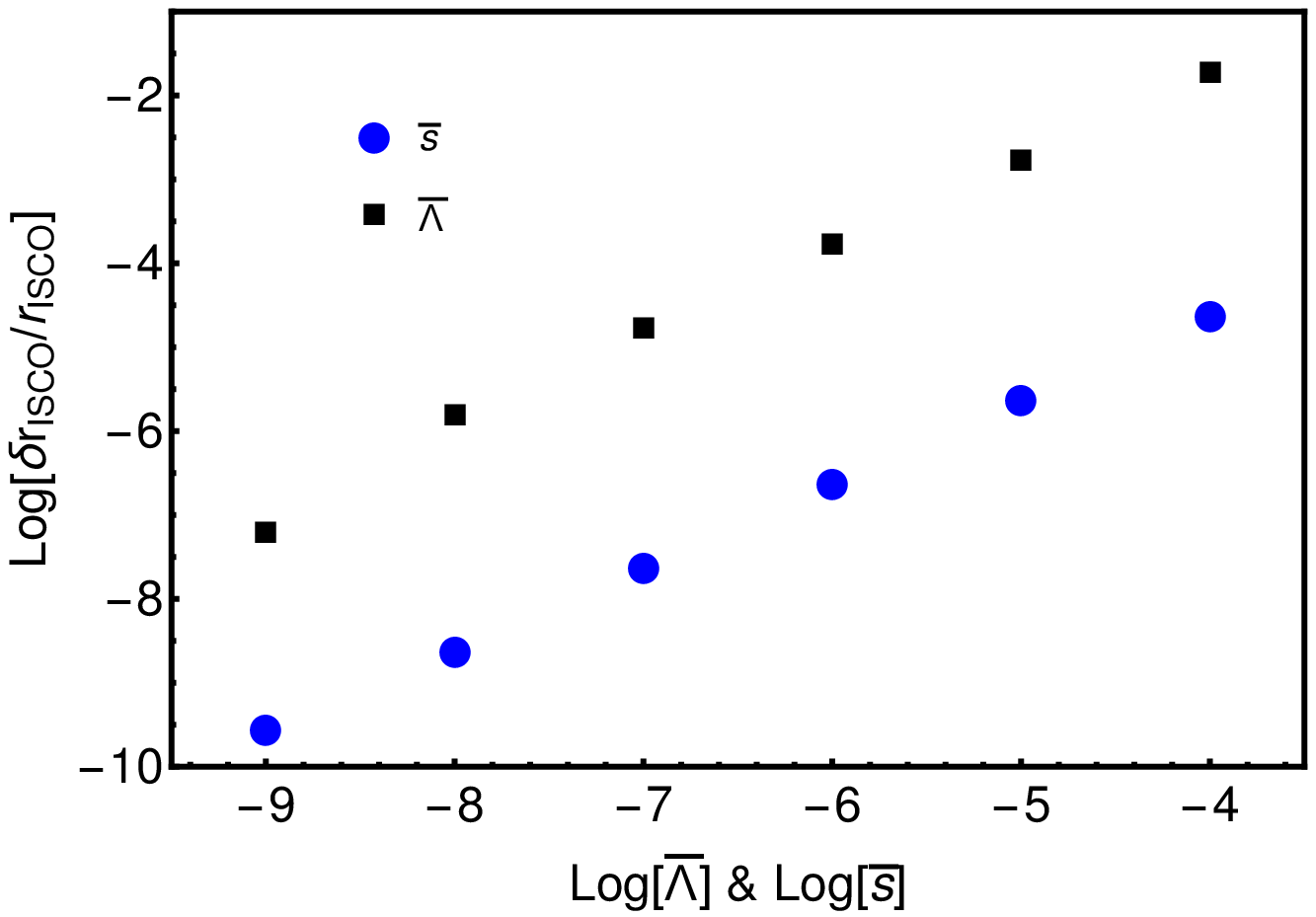}}
	\subfigure[~$a=0.25$(co-rotating orbit)]{\includegraphics[width=0.3\textwidth]{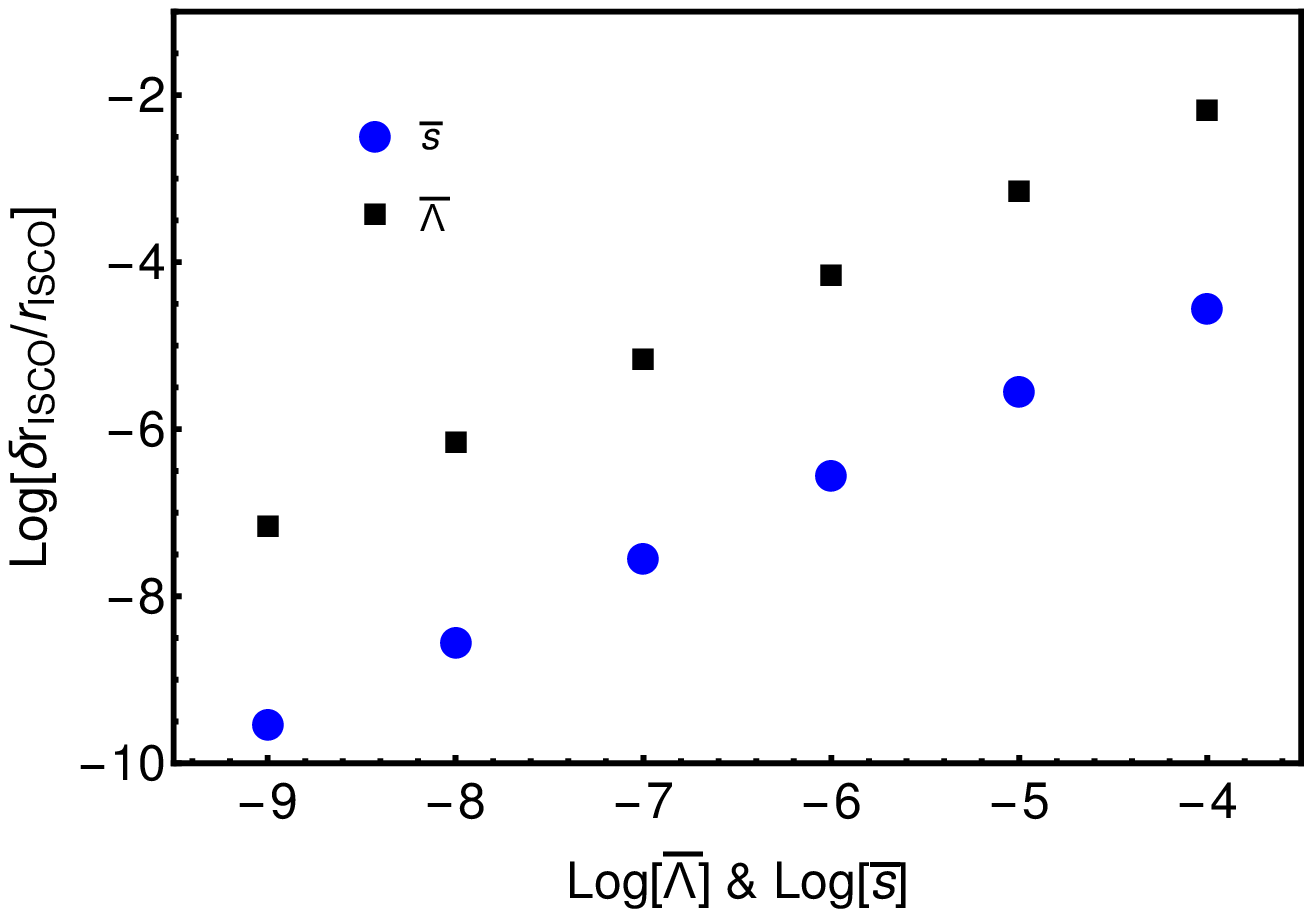}}
	
	\subfigure[~$a=0$]{\includegraphics[width=0.3\textwidth]{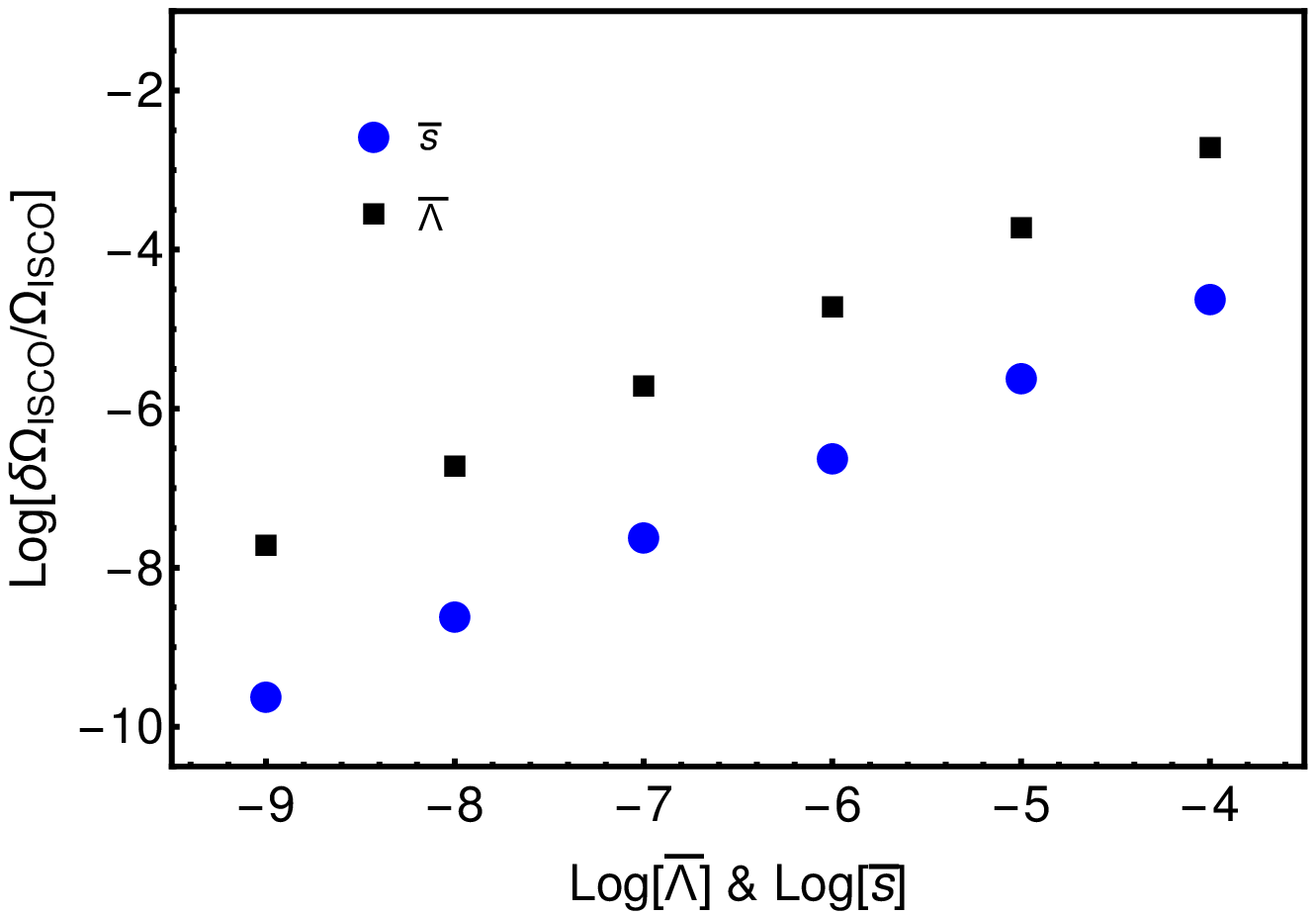}}
	\subfigure[~$a=0.25$(counter-rotating orbit)]{\includegraphics[width=0.3\textwidth]{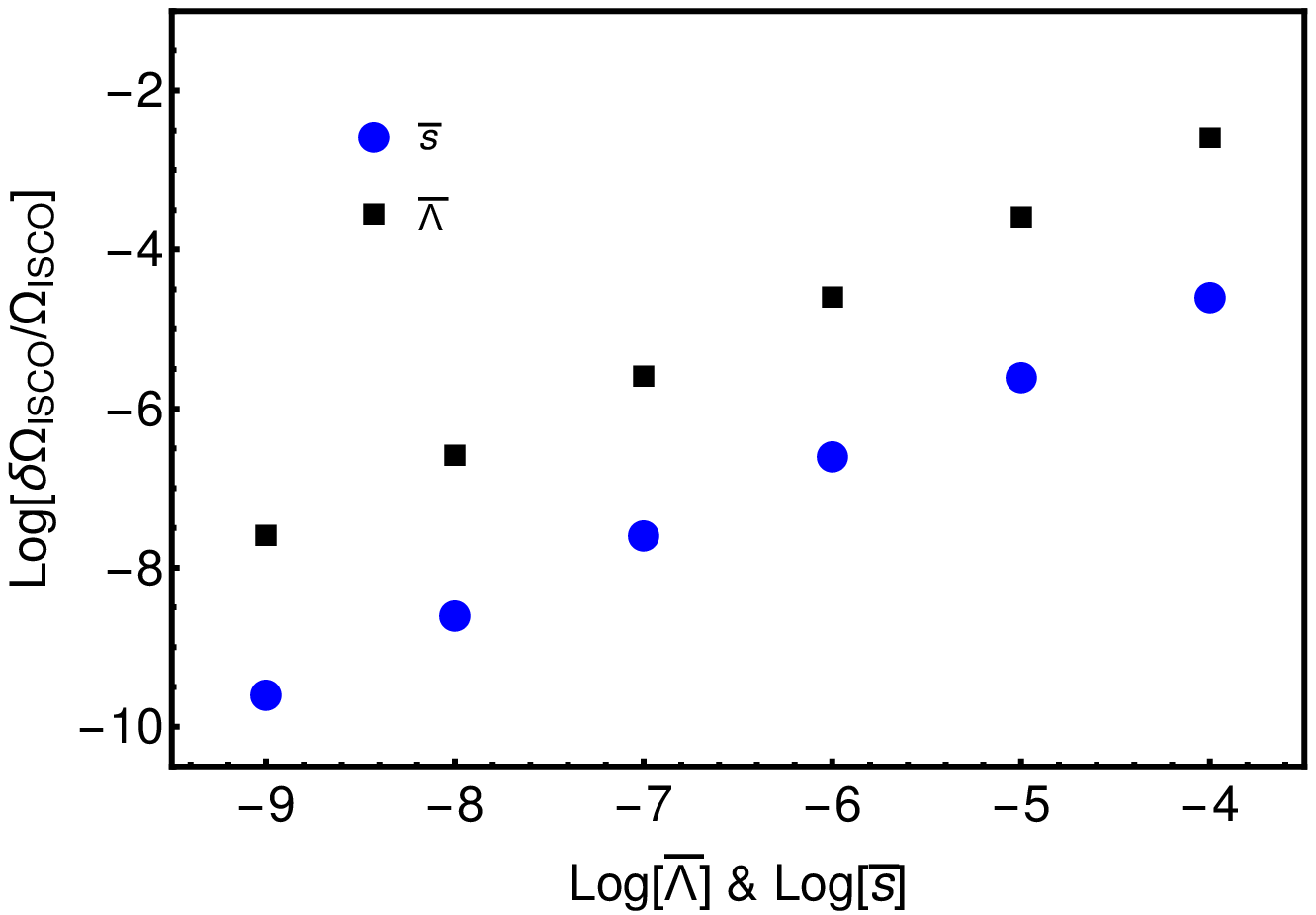}}
	\subfigure[~$a=0.25$(co-rotating orbit)]{\includegraphics[width=0.3\textwidth]{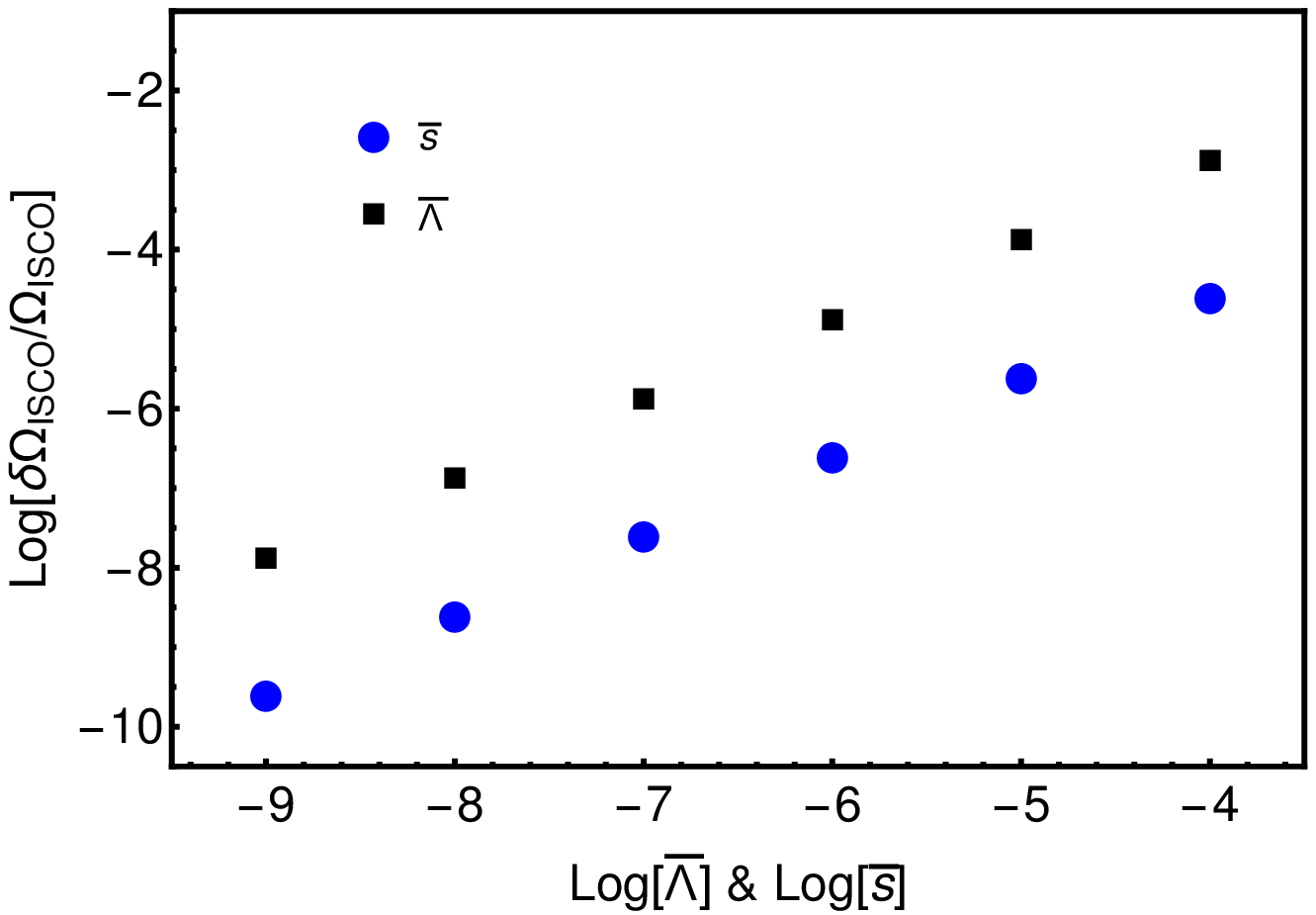}}
    \vskip -0mm \caption{Plots of the derivation of the ISCO parameters with different order of particle's spin and cosmological constant. The orders of the particle's spin and cosmological constant are set as $10^{-9}, 10^{-8}, 10^{-7}, 10^{-6}, 10^{-5},$ and $10^{-4}$.}
    \label{derivationofrandomega}
    \end{figure*}

Next, we will see that the relativistic values of the particle's spin $\bar{s}$ and cosmological constant $\bar{\Lambda}$ are very small and they are not of the same order. We define the following variables to describe the order of the correction induced by the particle's spin and cosmological constant as follows
\begin{eqnarray}
\delta r_{ \text{ISCO}, \bar{s}(\bar{\Lambda}) }|_{\bar{\Lambda}=0(\bar{s}=0)} &=& \frac{ r_{ \text{ISCO},\bar{s}(\bar{\Lambda}) } -r_{\text{ISCO} } }{r_{\text{ISCO}}},\\
\delta \Omega_{ \text{ISCO}, \bar{s}(\bar{\Lambda}) }|_{\bar{\Lambda}=0(\bar{s}=0)} &=& \frac{ \Omega_{ \text{ISCO},\bar{s}(\bar{\Lambda})  } -\Omega_{\text{ISCO} } }{\Omega_{\text{ISCO}}}.
\end{eqnarray}
The corresponding deviations for the motion of the test particle are shown in Fig. \ref{derivationofrandomega} and listed in Tables. \ref{ISCOparameters1}, \ref{ISCOparameters2}, \ref{ISCOparameters3}, \ref{ISCOparameters4}, \ref{ISCOparameters5}, and \ref{ISCOparameters6}. One thing should be noted here is that the same order of  $\bar{s}$ and $\bar{\Lambda}$ can make different order of corrections to the motion of the particle. The impact of the linear order parameter $\bar{\Lambda}$ is about two magnitudes greater than the one of the spin parameter $\bar{s}$ at the ISCO, we will discuss it lately.

\begin{table}[!htb]
\begin{center}
\caption{The ISCO parameters of the spinning test particle in the black hole background with $a=0$ and $\bar{\Lambda}=0$}.
\begin{tabular}{c c c c c c}
\hline
\hline
~~$\bar{s}$~~~&~~$\frac{r_{\text{ISCO}}}{M}$~~&~~$\bar{j}_{1\text{ISCO}}$~~&~~$\bar{e}_{\text{ISCO}}$~~\\
\hline
$1\times10^{-9}$ & 6.0000000019 & 3.4641016146 & 0.9428090416\\
$1\times10^{-8}$ & 6.0000000168 & 3.4641016104 & 0.9428090417\\
$1\times10^{-7}$ & 6.0000001636 & 3.4641015679 & 0.9428090432\\
$1\times10^{-6}$ & 6.0000016334 & 3.4641011437 & 0.9428090576\\
$1\times10^{-5}$ & 6.0000163303 & 3.4640969010 & 0.9428092020\\
$1\times10^{-4}$ & 6.0001633001 & 3.4640544744 & 0.9428106453\\
\hline
\hline
\end{tabular}
\label{ISCOparameters1}
\end{center}
\end{table}

\begin{table}[!htb]
\begin{center}
\caption{The ISCO parameters of the spinning test particle in the black hole background with $a=0$ and $\bar{s}=0$}.
\begin{tabular}{c c c c c c}
\hline
\hline
~~$\bar{\Lambda}$~~~&~~$\frac{r_{\text{ISCO}}}{M}$~~&~~$\bar{j}_{1\text{ISCO}}$~~&~~$\bar{e}_{\text{ISCO}}$~~\\
\hline
$1\times10^{-9}$ & 6.0000006484  &     3.4641014903    &    0.9428090246\\
$1\times10^{-8}$ & 6.0000064802  &     3.4641003679    &    0.9428088719\\
$1\times10^{-7}$ & 6.0000648048  &     3.4640891442    &    0.9428073445\\
$1\times10^{-6}$ & 6.0006484677  &     3.4639769053    &    0.9427920710\\
$1\times10^{-5}$ & 6.0065271010  &     3.4628543148    &    0.9426393359\\
$1\times10^{-4}$ & 6.0692859436  &     3.4514357844    &    0.9411003230\\
\hline
\hline

\end{tabular}
\label{ISCOparameters2}
\end{center}
\end{table}

\begin{table}[!htb]
\begin{center}
\caption{The ISCO (counter-rotating orbit) parameters of the spinning test particle in the black hole background with $a=0.25$ and $\bar{\Lambda}=0$}.
\begin{tabular}{c c c c c c}
\hline
\hline
~~$\bar{s}$~~~&~~$\frac{r_{\text{ISCO}}}{M}$~~&~~$\bar{j}_{1\text{ISCO}}$~~&~~$\bar{e}_{\text{ISCO}}$~~\\
\hline
$1\times10^{-9}$ & 6.7948537709   &    -3.6855902359    &    0.9496770967\\
$1\times10^{-8}$ & 6.7948537859   &    -3.6855902313    &    0.9496770968\\
$1\times10^{-7}$ & 6.7948539372   &    -3.6855901854    &    0.9496770980\\
$1\times10^{-6}$ & 6.7948554486   &    -3.6855897260    &    0.9496771092\\
$1\times10^{-5}$ & 6.7948705614   &    -3.6855851315    &    0.9496772214\\
$1\times10^{-4}$ & 6.7950216919   &    -3.6855391874    &    0.9496783434\\
\hline
\hline

\end{tabular}
\label{ISCOparameters3}
\end{center}
\end{table}

\begin{table}[!htb]
\begin{center}
\caption{The ISCO (counter-rotating orbit) parameters of the spinning test particle in the black hole background with $a=0.25$ and $\bar{s}=0$}.
\begin{tabular}{c c c c c c}
\hline
\hline
~~$\bar{\Lambda}$~~~&~~$\frac{r_{\text{ISCO}}}{M}$~~&~~$\bar{j}_{1\text{ISCO}}$~~&~~$\bar{e}_{\text{ISCO}}$~~\\
\hline
$1\times10^{-9}$ & 6.7948554128  &  -3.6855900062 & 0.9496770728\\
$1\times10^{-8}$ & 6.7948651402  &  -3.6855882896 & 0.9496768779\\
$1\times10^{-7}$ & 6.7949626058  &  -3.6855713142 & 0.9496749402\\
$1\times10^{-6}$ & 6.7959416478  &  -3.6854025138 & 0.9496556180\\
$1\times10^{-5}$ & 6.8058299155  &  -3.6837127936 & 0.9494623240\\
$1\times10^{-4}$ & 6.9147050109  &  -3.6663300559 & 0.9475046091\\
\hline
\hline

\end{tabular}
\label{ISCOparameters4}
\end{center}
\end{table}

\begin{table}[!htb]
\begin{center}
\caption{The ISCO (co-rotating orbit) parameters of the spinning test particle in the black hole background with $a=0.25$ and $\bar{\Lambda}=0$}.
\begin{tabular}{c c c c c c}
\hline
\hline
~~$\bar{s}$~~~&~~$\frac{r_{\text{ISCO}}}{M}$~~&~~$\bar{j}_{1\text{ISCO}}$~~&~~$\bar{e}_{\text{ISCO}}$~~\\
\hline
$1\times10^{-9}$ & 5.1559328132   &    3.2099521130   &     0.9331177083\\
$1\times10^{-8}$ & 5.1559328269   &    3.2099521091   &     0.9331177084\\
$1\times10^{-7}$ & 5.1559329658   &    3.2099520706   &     0.9331177103\\
$1\times10^{-6}$ & 5.1559343514   &    3.2099516854   &     0.9331177292\\
$1\times10^{-5}$ & 5.1559482101   &    3.2099478341   &     0.9331179181\\
$1\times10^{-4}$ & 5.1560867956   &    3.2099093195   &     0.9331198064\\
\hline
\hline

\end{tabular}
\label{ISCOparameters5}
\end{center}
\end{table}

\begin{table}[!htb]
\begin{center}
\caption{The ISCO (co-rotating orbit) parameters of the spinning test particle in the black hole background with $a=0.25$ and $\bar{s}=0$}.
\begin{tabular}{c c c c c c}
\hline
\hline
~~$\bar{\Lambda}$~~~&~~$\frac{r_{\text{ISCO}}}{M}$~~&~~$\bar{j}_{1\text{ISCO}}$~~&~~$\bar{e}_{\text{ISCO}}$~~\\
\hline
$1\times10^{-9}$ & 5.1559331501    &   3.2099520350   &     0.9331176953\\
$1\times10^{-8}$ & 5.1559361936    &   3.2099513292   &     0.9331175786\\
$1\times10^{-7}$ & 5.1559666327    &   3.2099442709   &     0.9331164123\\
$1\times10^{-6}$ & 5.1562711583    &   3.2098736874   &     0.9331047488\\
$1\times10^{-5}$ & 5.1593301318    &   3.2091677670   &     0.9329881144\\
$1\times10^{-4}$ & 5.1913776247    &   3.2020998180   &     0.9318217663\\
\hline
\hline

\end{tabular}
\label{ISCOparameters6}
\end{center}
\end{table}

Next, we will take the examples of kerr black hole and neutron star as the small body to estimate the magnitudes of the motion deviation induced by their spin. The spin angular momentum of a compact object is always referred by the so-called ``spin-parameter''
\begin{equation}
b=\frac{cs}{Gm ^2}=\frac{s}{m^2},
\end{equation}
where the parameters $s$ and $m$ are the angular momentum and mass of the compact object. The mass of the central supermassive black hole in this system is assumed to be $10^6 M_\odot$, where the parameter $M_\odot$ is the solar mass. The spin angular momentum of the test body satisfies $s=bm^2$, and the dimensionless spin parameter reads
\begin{equation}
\bar{s}=\frac{bm^2}{mM}=b\frac{m}{M}\ll1.
\end{equation}
If we take the neutron star as the small body, and choose the mass and spin parameter as $m_{\text{NS}}=1.4M_\odot$ and $b_{\text{NS}}\sim 0.02$ \cite{Miller:2011jq}, then the spin parameter $\bar{s}$ for the neutron star is $\bar{s}=b\frac{m}{M}\sim 10^{-8}$. Even for a small extremal Kerr black hole with mass $m_{\text{ek}}\sim10 M_\odot$, the corresponding spin parameter $\bar{s}=\frac{m}{M}\sim 10^{-5}$ is still very small.

The cosmological constant is given by
\begin{equation}
\Lambda=3\left(\frac{H_0}{c}\right)^2 \Omega_\Lambda,
\end{equation}
where the parameters $H_0$, $c$, and $\Omega_\Lambda$ are the Hubble constant, speed of light, and dark energy density parameter today. The current observational values of these parameters from Planck collaboration are \cite{Planck2018} (see Table A.1)
\begin{eqnarray}
H_0&=&67.66\pm0.42~\text{km}~\text{s}^{-1} \text{Mpc}^{-1},\\
\Omega_\Lambda&=&0.6897\pm 0.0057.
\end{eqnarray}
With these results, the value of cosmological constant is
\begin{equation}
\Lambda\sim 1.10536\times 10^{-52}m^{-2}.
\end{equation}
Then the corresponding dimensionless cosmological constant is
\begin{equation}
\bar{\Lambda}\sim 1.10536\times 10^{-34},
\end{equation}
which is very tiny. Now, we can have a quick glance at the dimensionless spin $\bar{s}$ and cosmological constant $\bar{\Lambda}$, which are so tiny and seemingly can be ignored. On the other hand, the phase shift might be considerable due to the accumulation during the inspiral. To check
this, we will analyze the order of phase shift due to the accumulation during the inspiral. Since the spin parameter $\bar{s}$ and cosmological constant $\bar{\Lambda}$ are so tiny, we can decompose the frequency $\Omega$ with the linear order of the particle's spin and cosmological constant as follows
\begin{eqnarray}
\Omega&=&\Omega_0+\delta\Omega_1+\delta\Omega_2\nonumber\\
&=&\Omega_0+\phi_s~\bar{s}+\phi_\Lambda~\bar{\Lambda}+\big(\mathcal{O}(\bar{s})+\mathcal{O}(\bar{\Lambda})\big)^2,
\label{angular-frequency-all}
\end{eqnarray}
where $\delta\Omega_1=\phi_{s}\bar{s}$, $\delta\Omega_2=\phi_{\Lambda}\bar{\Lambda}$, and $\Omega_0$ is the angular frequency of the spinless test particle in the Kerr black hole background with zero cosmological constant. The corresponding complete expressions are
\begin{eqnarray}
\Omega_0&=&\frac{2 a \bar{e} M-2 \bar{j} M^2+\bar{j} M r}{2 a^2 \bar{e} M+a^2 \bar{e} r-2 a \bar{j} M^2+\bar{e} r^3},
\label{angular-frequency-0}\\
\phi_s&=&\frac{a^2 \!+\!r (r\!-\!2 M)}{r}\Big[2 a^2 \bar{e} M\!+\!a^2 \bar{e} r\!-\!2 a \bar{j} M^2 \!+\!\bar{e} r^3\Big]^{-2}\nonumber\\
&&\times\Big\{-M\big[M(a\bar{e}-\bar{j}M)^2+a^2r(\bar{e}^2-1)\nonumber\\
&&-\bar{j}^2M^2r+2Mr^2+(2\bar{e}^2-1)r^3\big]\Big\},
\label{frequency-shift-from-s} \\
\phi_\Lambda&=&\frac{1}{3M^2}\left[2 a^2 \bar{e} M+a^2 \bar{e} r-2 a \bar{j} M^2+\bar{e} r^3\right]^{-2}\nonumber\\
&&\times\Big\{a^3\big[2\bar{e}^2r^3(M+r)+\bar{j}^2M^2(4M^2-2Mr\nonumber\\
&&+r^2)\big]-8a^4\bar{e}\bar{j}M^3-\bar{e}\bar{j}Mr^6\nonumber\\
&&-a^2\bar{e}\bar{j}M r^3(4M+r)+a^5\bar{e}^2(4M^2\nonumber\\
&&+2Mr+r^2)+a(\bar{j}^2M^2r^4+\bar{e}^2r^6)\Big\}.
\label{frequency-shift-from-lambda}
\end{eqnarray}
Note that, the linear order shift of the angular frequency (\ref{frequency-shift-from-s}) induced by the particle's spin will vanish when the orbit locates at the horizon. This behavior is consistent with the results in Ref. \cite{Jefremov:2015gza}.

Now we can use Eqs. (\ref{frequency-shift-from-s}) and (\ref{frequency-shift-from-lambda}) to estimate the frequency shift induced by the cosmological constant and particle's spin. Let us take a cycle of the ISCO as an example to compute the phase shift over a period. We cursorily choose the period of the ISCO as $T=\frac{2\pi}{\Omega_0}$ and list the quantities $\Omega_0$, $\phi_\Lambda$, $\phi_s$ given in Eqs. (\ref{angular-frequency-all})-(\ref{frequency-shift-from-s}) with different values of the spinning black hole for the ISCO in Table. \ref{frequency-shift}. Then the corresponding accumulations of the phase shift induced by the particle's spin and cosmological constant over a whole period can be calculated as
    \begin{eqnarray}
        \delta\Omega^{\text{1\,period}}_{1}=T \delta\Omega_1& \sim&\left\{\begin{array}{cl}
                                  5.0\times 10^{-7}\pi, & ~~~~~a=-0.2,\\
                                  5.1\times 10^{-7}\pi, & ~~~~~a=0, \\
                                  5.2\times 10^{-7}\pi, & ~~~~~a=0.2,\\

                                \end{array} \right.\\
        \delta\Omega^{\text{1\,period}}_{2}= T \delta\Omega_2 &\sim &\left\{\begin{array}{cl}
                                   -4.6\times 10^{-33}\pi, & ~a=-0.2, \\
                                   -3.6\times 10^{-33}\pi, & ~a=0, \\
                                   -2.7\times 10^{-33}\pi, & ~a=0.2, \\

                                \end{array} \right.
    \end{eqnarray}
where the parameters are set as $\bar{s}=10^{-6}$ and $\bar{\Lambda}=10^{-34}$.
As the results in Ref. \cite{thorne2000}, for a $m=10M_\odot$(small black hole) test body spiraling into a black hole with mass $M=10^6M_\odot$ from $r=9.46 M$ to the ISCO at $r=6M$ with circular orbits, the motion of the test body will be about 85000 cycles. Then the corresponding magnitudes of the
phase shift $\delta\Omega_1$ and $\delta\Omega_2$ induced by the particle's spin
and cosmological constant in this region can be roughly computed by setting the same angular frequency shift for each period. They are about
    \begin{eqnarray}
        \delta\Omega_{1}&\sim& 10^{-2}\pi,\\
        \delta\Omega_{2}&\sim& 10^{-28}\pi.
    \end{eqnarray}
For an entire inspiraling process from more further distance to the ISCO, the phase shift induced by the particle's spin will be considerable. While the phase shift induced by the cosmological constant is actually tiny compared the one induced by the particle's spin.

\begin{table}[!htb]
\begin{center}
\caption{The quantities $\Omega_0$, $\phi_\Lambda$, $\phi_s$ in Eqs. (\ref{angular-frequency-all})-(\ref{frequency-shift-from-s}) with different values of the spinning black hole at the ISCO.}
\begin{tabular}{c c c c}
\hline
\hline
~~~~~$a$~~~~~       &~~~~$\Omega_0$~~~~  &~~~~$\phi_\Lambda$~~~~ &~~~~$\phi_s$~~~~\\
\hline
 -0.2 &  $\frac{0.059144}{M}$ & $-\frac{1.35394}{M}$ & $-\frac{0.01480}{M}$\\
\hline
 0 & $\frac{1}{6\sqrt{6}M}$ & $-\sqrt{\frac{3}{2}}\frac{1}{M}$ & $-\frac{5}{288M}$\\
\hline
 0.2 &  $\frac{0.079974}{M}$ & $-\frac{1.08631}{M}$ & $-\frac{0.02077}{M}$\\
\hline
\hline
\end{tabular}\\
\label{frequency-shift}
\end{center}
\end{table}

Before closing this section, we make a brief discussion about our results. As the results of the motion deviation induced by particle's spin and cosmological constant in Fig. \ref{derivationofrandomega}, the same order of cosmological constant $\bar{\Lambda}$ and particle's spin $\bar{s}$ can induce different magnitudes of deviation to the motion of the test particle. To explain the reason for this phenomenon, we can use the angular frequency (\ref{angular-frequency-all}) with linear particle's spin $\bar{s}$ and cosmological constant $\bar{\Lambda}$ to compute the contributions induced by the particle's spin and cosmological constant. Then the the ratio of corrections induced between cosmological constant and particle at ISCO is
\begin{eqnarray}
\frac{\phi_\Lambda}{\phi_s}\sim 10^2,
\end{eqnarray}
where the results in Table. \ref{frequency-shift} have been used. On the other hand, the inclination of the plots in Fig. \ref{derivationofrandomega} is still related to the linear order functions $\phi_s$ and $\phi_\Lambda$, actually the slope of the curves can be determined by using $\frac{\phi_\Lambda}{\Omega_0}$ and $\frac{\phi_s}{\Omega_0}$. We should note that the impacts of the perturbation parameters depend on the radius, and the corrections induced by the cosmological constant is two magnitudes larger than the one induced by the same order particle's spin does not hold everywhere. Next, we will give the relations of the corrections induced by between the particle's spin and the cosmological constant by considering circular orbits. The angular frequency of the test particle at circular orbit with vanishing particle's spin and cosmological constant reads \cite{Bardeen1972}
\begin{equation}
\Omega_0=\frac{M^{1/2}}{r^{3/2}+ a M^{1/2}}.
\label{circularangularfrequency}
\end{equation}

\begin{figure*}[!htb]
	\includegraphics[width=0.3\textwidth]{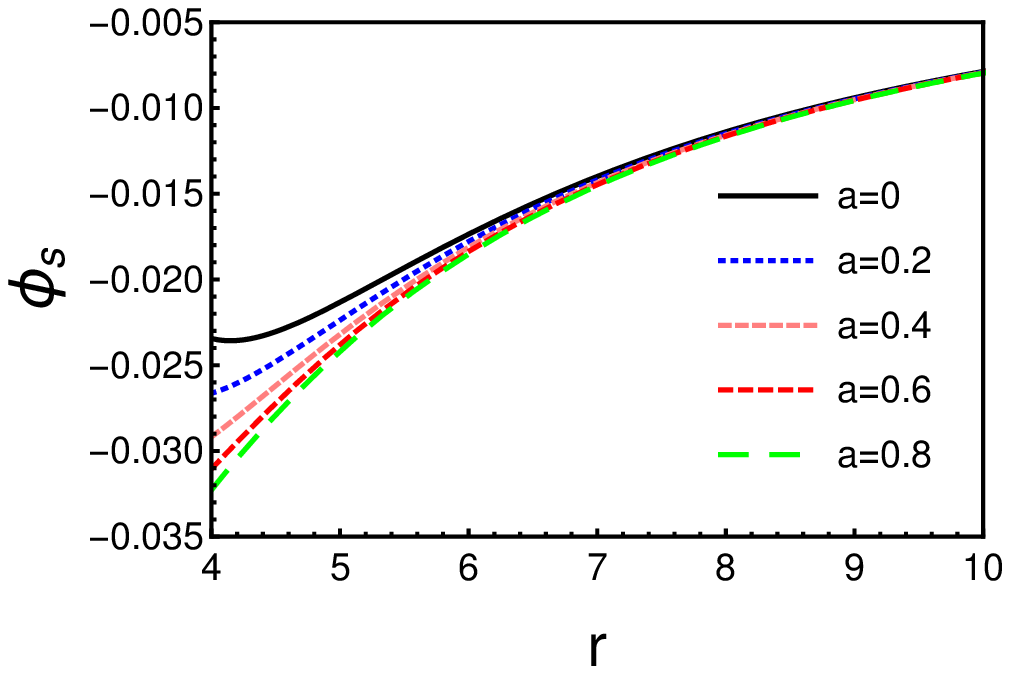}
	\includegraphics[width=0.3\textwidth]{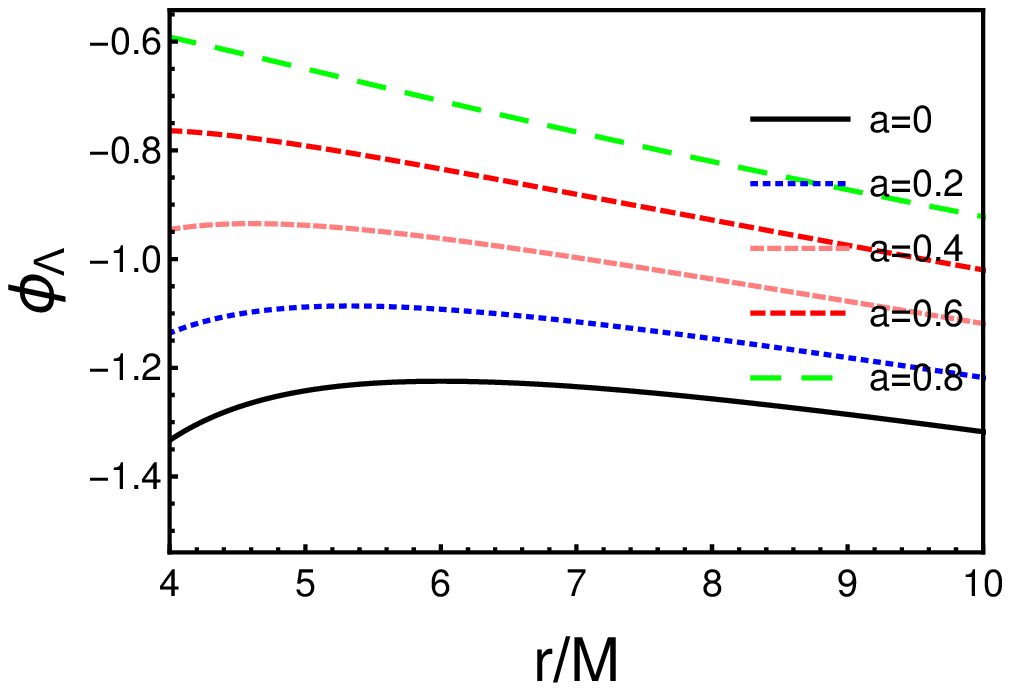}
	\includegraphics[width=0.3\textwidth]{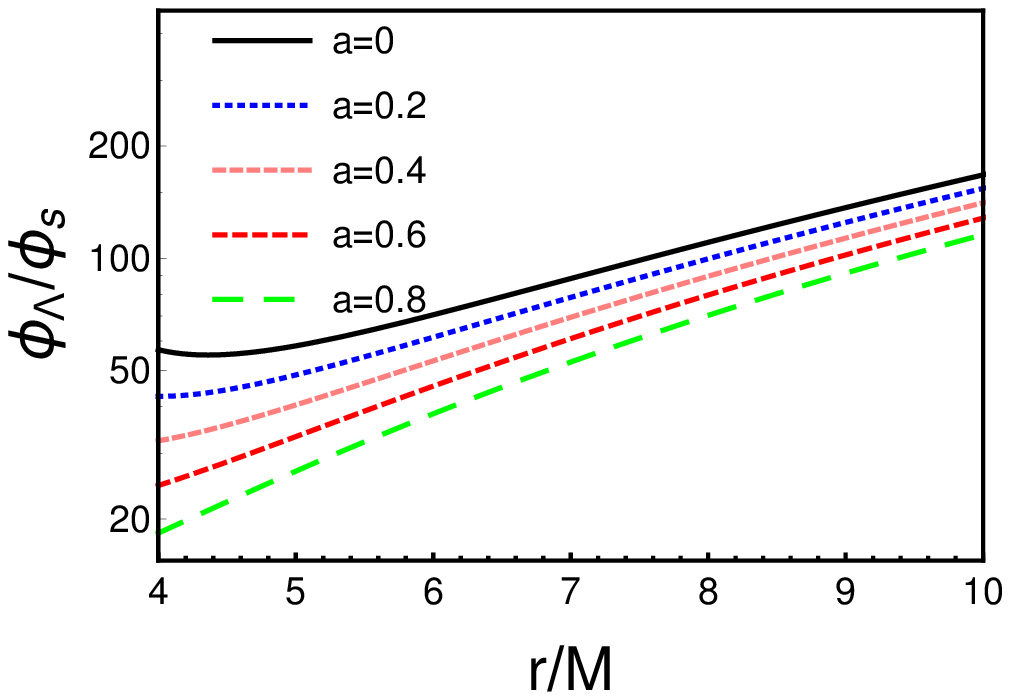}

	\includegraphics[width=0.3\textwidth]{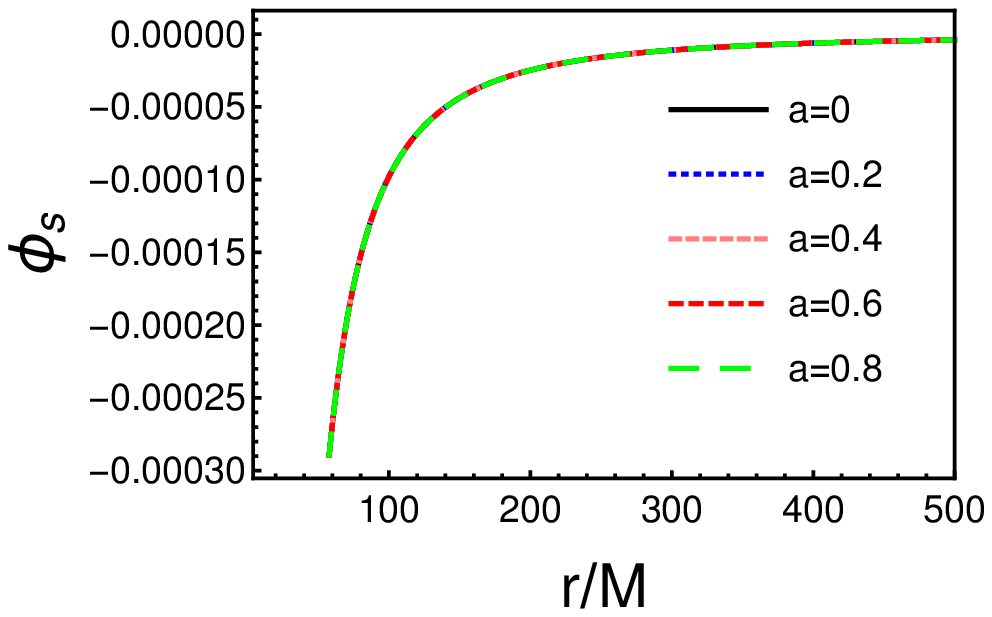}
	\includegraphics[width=0.3\textwidth]{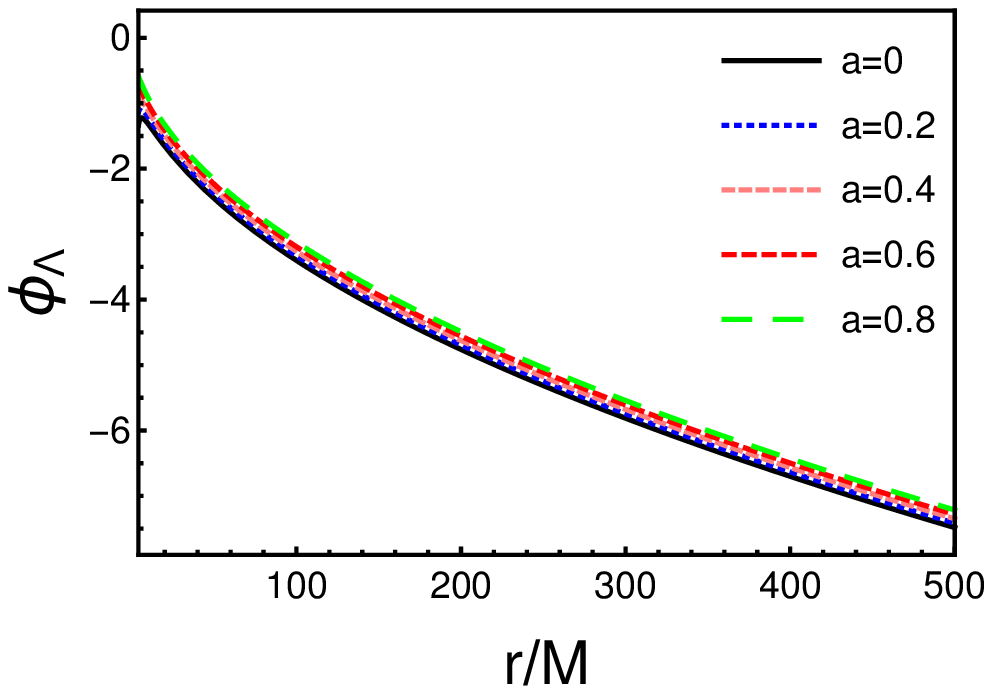}
	\includegraphics[width=0.3\textwidth]{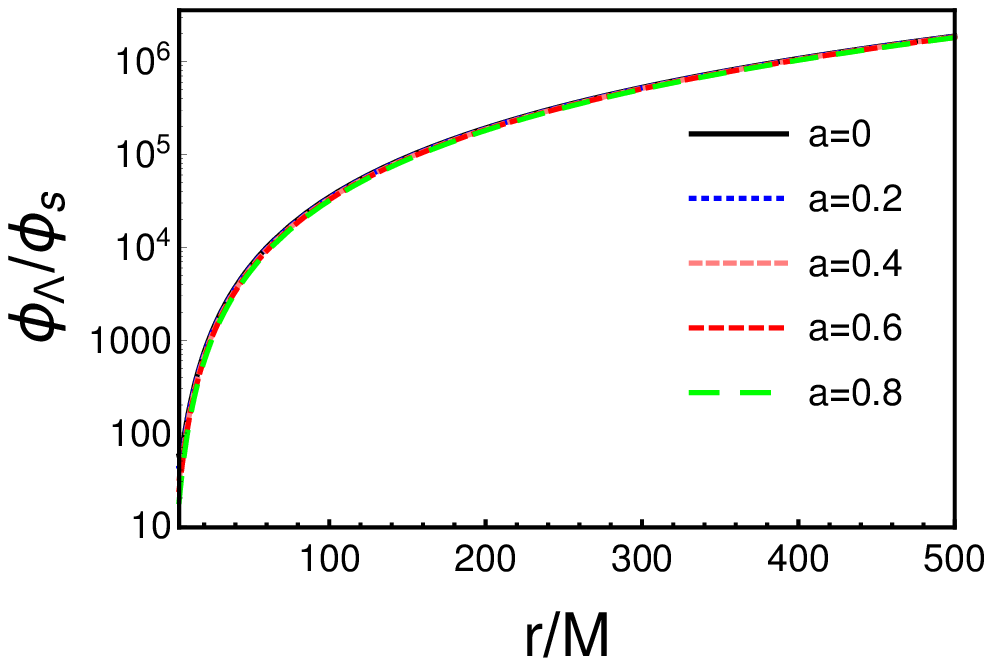}	
    \vskip -0mm \caption{Plots of the derivation angular frequency for circular orbit induced by particle's spin and cosmological constant that how they depend on the radius. The top channel for the details in small radius range, the bottom channel for the details in large radius range.}
    \label{derivationofrandomegaradius}
    \end{figure*}

Then by substituting the energy and angular momentum in circular orbit into Eqs. (\ref{frequency-shift-from-s}) and (\ref{frequency-shift-from-lambda}), we plot the corrections for the angular frequency within a linear order approximation and show how they depend on the radius, see Fig. \ref{derivationofrandomegaradius}.

Obviously, for the whole process of the inspiral, the accumulated phase shift needs to be handled with caution in terms of the relations in Fig. \ref{derivationofrandomegaradius}. Although we have shown that the current observed value of the dimensionless cosmological constant is much smaller than the value of dimensionless particle's spin, however the impact of the cosmological constant will be much greater than the one of the same order spin when the particle is located far away from the black hole. It is easy to see that the ratio $\frac{\phi_\Lambda}{\phi_S}$ is no longer sensitive to the spin of the black hole when the radius becomes bigger and we will ignore the impacts of the black hole's spin. By using Eqs. (\ref{frequency-shift-from-s}) and (\ref{frequency-shift-from-lambda}), the dependence of motion's deviation, induced by the spin and the cosmological constant, on the radius can be cursorily obtained as follows
\begin{eqnarray}
\phi_\Lambda|_{a=0} &=&-\frac{\sqrt{ r^3}}{3 (r-2)},\\
\phi_s|_{a=0} &=&- \frac{(r-1) (r-3)}{(r-2) r^3},
\end{eqnarray}
then the corresponding ratio is
\begin{equation}
\frac{\phi_\Lambda}{\phi_s}|_{a=0}=\frac{r^5}{3 (r-3) (r-1) \sqrt{r}}.
\end{equation}
An interesting phenomenon exists here, the ratio will diverge when the radius approaches to infinity. The reason for this behavior is mainly from that the ``spin-curvature" coupling will decraease with the radius $r$, while the impact that from the cosmological constant is opposite.

Finally, we can have a look at again for the phase shift induced by particle's spin and cosmological constant. The angular frequency can be estimated by using Eq. (\ref{circularangularfrequency}), if we naively consider the detectable frequency range for LISA as
\begin{equation}
\Omega_0\in (10^{-4}\text{Hz},10^{-1}\text{Hz}),
\end{equation}
and still set the mass of supermassive black hole as $10^6M_\odot$. Then the corresponding radius range of the circular orbirt is
\begin{equation}
r\in (10^0M,10^2M),
\label{radiusrange}
\end{equation}
where $M$ is the mass of the supermassive black hole.

Although we have shown that the contribution induced by cosmological constant will increase with radius, however when the test particle locates at the range of Eq. (\ref{radiusrange}), the contribution from the particle's spin is still far greater than the one from the cosmological constant.

\section{Summary and conclusion}\label{Conclusion}

In this paper, we investigated the motion deviation of the small body induced by the particle's spin and the non-zero cosmological constant. To derive the corrections  induced by the particle's spin and the cosmological constants with a unified treatment, we solved the equation of motion for a spinning test particle by using the MPD equation with Tulczyjew spin-supplementary condition in the equatorial plane of the Kerr-dS/AdS black hole background, and gave the four-momentum and four-velocity of the spinning test particle. Since the radial component of the tangent vector $\dot{r}$ and the radial component of the four-momentum $P^r$ are parallel, we derived the effective potential by decomposing the radial component of the four-momentum and used it to obtain the ISCO of the spinning test particle.

By using the ISCO of the test particle, we estimated how they are changed by the non-vanishing particle's spin and cosmological constant. We numerically investigated the corrections for the ISCO that from the particle's spin and cosmological constant with small order values, we found that the same order particle's spin and cosmological constant can make different order of contributions to the motion of the test particle. By considering a small Kerr black hole or a rotating neutron star as the small test spinning body and considering the current observations of the cosmological constant, we got two different order of magnitudes of parameters $\bar{s}={s}/{(mM)}$ and $\bar{\Lambda}=\Lambda~M^2$. The linear order analytic corrections to the angular frequency induced by the particle's spin and cosmological constant in the background of Kerr black hole with arbitrary black hole spin $a$ were given, and it will work well due to the so tiny true values of the particle's spin and cosmological constant. With our linear order correction (\ref{angular-frequency-all}), we estimated the magnitudes of phase shift induced by the particle's spin and cosmological constant over a period at the ISCO. We also showed that ratio of the impacts that from the particle's spin and cosmological constant does not hold everywhere and depends on the location of the orbit. The analytic ratio of the contribution induced by between the cosmological constant and the particle's spin was still given. It will be useful for the more accurate phase shift of the whole inspiraling process for the dynamical EMRI system with gravitational emission.

\section{Acknowledgments}

We thank the anonymous referee's useful suggestions for this paper, which played a great role to the improvement of this paper. This work was supported in part by the National Natural Science Foundation of
China (Grants No. 11875151, No. 11675064, and No. 11522541), the Strategic Priority Research Program on Space Science, the Chinese Academy of Sciences (Grant No. XDA15020701), and the
Fundamental Research Funds for the Central Universities (Grants No.
lzujbky-2019-ct06, No. lzujbky-2018-k11,  and No. lzujbky-2017-it68). Y.P. Zhang was supported by the scholarship granted by the Chinese Scholarship Council (CSC).
PAS acknowledges support from the Ram{\'o}ny Cajal Programme of the Ministry
of Economy, Industry and Competitiveness of Spain, as well as the COST Action
GWverse CA16104. This work was supported by the National Key R\&D Program of
China (2016YFA0400702) and the National Science Foundation of China (11721303).


\end{document}